\documentclass{jaa}
\usepackage{natbib}
\usepackage{xcolor}
\usepackage{hyperref}
\usepackage{amsmath}
\usepackage{amssymb}
\usepackage{aas_macros}
\usepackage{graphicx}

\hypersetup{
    colorlinks=true,
    linkcolor=red,
    citecolor=blue,
    filecolor=magenta, 
    urlcolor=blue,
}

\begin{document}\sloppy

\title{Pulsar timing irregularities and the Neutron Star interior in the era of SKA: An Indian Outlook}


\author{Jaikhomba Singha\textsuperscript{1,*}, Bhal Chandra Joshi\textsuperscript{2},  Debades Bandyopadhyay\textsuperscript{3}, Himanshu Grover\textsuperscript{1},\\ Shantanu Desai\textsuperscript{4}, P. Arumugam\textsuperscript{1} and Sarmistha Banik\textsuperscript{5}}
\affilOne{\textsuperscript{1}Department of Physics, Indian Institute of Technology Roorkee, Roorkee 247667, India.\\}
\affilTwo{\textsuperscript{2}National Centre for Radio Astrophysics - Tata Insitute of Fundamental Research, Pune$-$411007, India\\}
\affilThree{\textsuperscript{3}Saha Institute of Nuclear Physics, Kolkata$-$700064, India.\\}
\affilFour{\textsuperscript{4}Department of Physics, Indian Institute of Technology Hyderabad, Kandi, Telangana 502285, India\\}
\affilFive{\textsuperscript{5}Department of Physics, BITS Pilani, Hyderabad Campus, Hyderabad 500078, Telangana, India.\\}


\twocolumn[{

\maketitle

\corres{mjaikhomba@gmail.com}

\msinfo{XXX}{YYY}

\begin{abstract}
There are two types of timing irregularities seen in pulsars: glitches and timing noise. Both of these phenomena can help us probe the interior of such exotic objects. This article presents a brief overview of the observational and theoretical aspects of pulsar timing irregularities and the main results from the investigations of these phenomena in India. The relevance of such Indian programs for monitoring of young pulsars with the Square Kilometer Array (SKA) is presented, highlighting possible contributions of the Indian neutron star community to the upcoming SKA endeavour.
\end{abstract}

}]

\doinum{12.3456/s78910-011-012-3}
\artcitid{\#\#\#\#}
\volnum{000}
\year{0000}
\pgrange{1--}
\setcounter{page}{1}
\lp{1}

\begin{figure*}[!ht]
\includegraphics[scale=0.7]{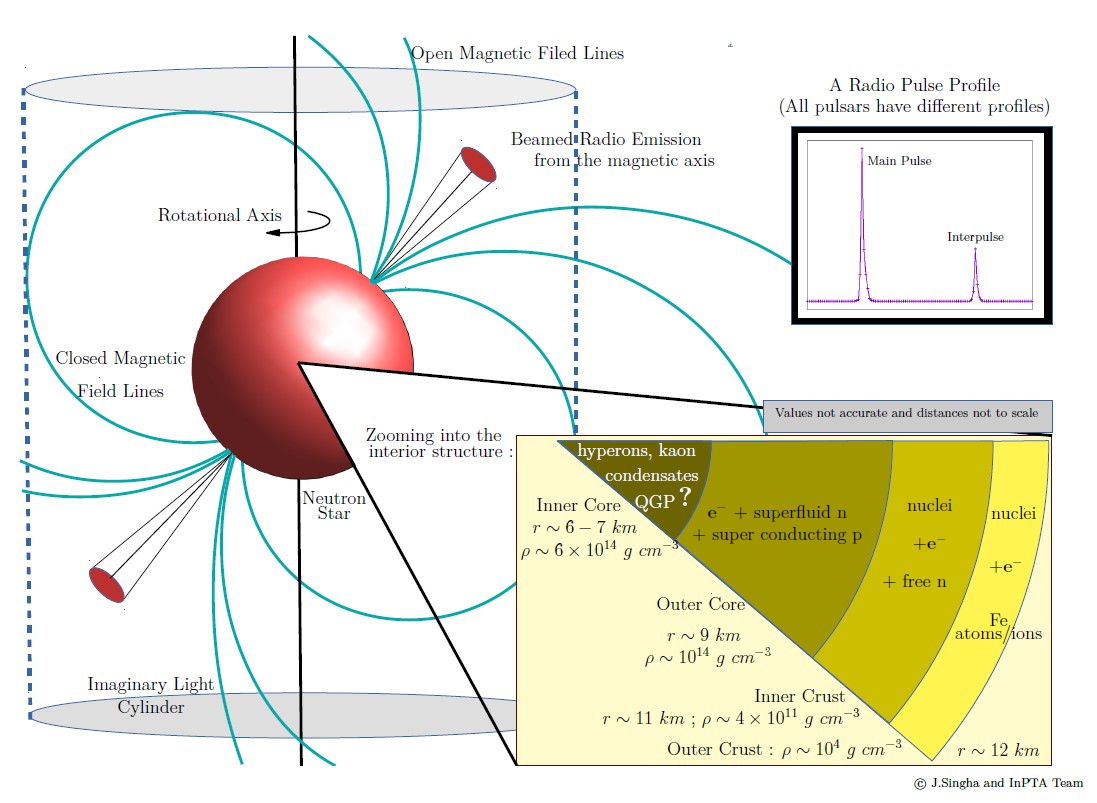}
\caption{An artist's impression of a Neutron star. The magnetosphere consists of both closed and open magnetic field lines. A beam of radio waves are emitted from the magnetic poles. In the bottom right, a zoomed-in map of the NS interior is shown. The interior consists of several layers with different compositions. An integrated pulse profile is shown in the top right. It is a stable pulse profile after being averaged over few thousands of periods.}\label{fig1}
\end{figure*}

\section{Introduction}
\label{Intro}
Neutron Stars (NSs) are the stellar remnants of massive stars formed as a result of supernova explosions~\citep{Baade}. These are very compact objects and exhibit extreme properties, which make them good laboratories to probe a variety of physics \citep{weber1999pulsars, Bandyopadhyay2017}. Radio pulsars are the observational manifestations of NSs. Pulsars are magnetized NSs that appear to emit periodic short pulses of radio radiation with periods between 1.4 ms \citep{1.4mspulsar} and 23.5 s \citep{23.5spulsar}. Pulsars are known to exhibit remarkable stable periodicity. However, there is a steady slow-down of their rotation rates due to the loss of rotational energy. This steady slow-down is occasionally interrupted by two irregularities: glitches \citep{Radman69} and timing noise \citep{Boynton_1972, Cordes_Helfand1980, Hobbs_2006, Lyne+2013}. Glitches are the sudden rise in the rotational frequency of pulsars, and timing noise is the quasi-random wandering seen in the pulsar spin period. Glitches are the direct manifestations of the internal structure of a neutron star. The glitch rise time and the post-glitch evolution of the rotational frequency are important parameters and can have varied timescales \citep{espinoza2011, Yu2013}. Observations and modelling of such quantities can help us probe the superfluid dynamics inside the star \citep{haskell+2018, erbil2022, BaymEipstein1988, Sauls1989}. The intricate behaviour regarding the participation of the stellar core can also be understood from large glitches. The recently observed slow rise in the glitch of PSR J0534+2200 (Crab pulsar) by \cite{Shaw+2018, basu2020} is an important observation, which also helps in constraining the strength of the mutual friction coefficient \citep{haskell+2018} between the normal and the superfluid component. Glitches are usually seen in young pulsars, where timing noise is also very prominent. The theoretical understanding of timing noise is still not very clear even after several decades of its discovery. Studying both these phenomena require regular timing observations of a large sample of pulsars. Due to a large number of antennas available, wide frequency coverage with wide-band receivers and subarray capabilities, the SKA will provide a perfect opportunity to help us answer some of these questions. As there are significant number of Indian pulsar astronomers and neutron star theorists, their participation in such observations with the SKA will provide an excellent opportunity to involve the Indian neutron star community. Some members of this community are already engaged in similar programs with the upgraded Giant Metrewave Radio Telescope (uGMRT) \citep{ugmrt} and the Ooty Radio Telescope (ORT) \citep{swarup1971large}. It may be noted that the uGMRT is a pathfinder telescope to the SKA, while the ORT is a large collecting area telescope at SKA-low frequencies. 

The present article provides a brief review of the state-of-art in the monitoring of rotational irregularities in NSs outlining the current pathfinder programs with the uGMRT and the ORT. The possible science outcomes with a more sensitive telescope like the SKA have been highlighted. The article has been organised in the following manner. A brief review of pulsar timing irregularities and how they can be used to probe the interior of the NS is presented in Section \ref{Pulsar Timing Irregularities and the NS interior} The glitch monitoring programs using Indian telescopes have been described in Section~\ref{Glitch Monitoring Programs in India} We describe the upcoming SKA facility, their applications for pulsars timing observation, and the science outcomes in Section \ref{Pulsar Observations} The lessons learnt from the current pathfinder programs at the uGMRT and the ORT, which are helpful for the SKA program, have been discussed in Section~\ref{Lessons} Finally, a summary of the future of such pulsar timing programs and the possible contributions from the Indian NS star community is given in Section~\ref{Conclusion}

\section{Pulsar Timing Irregularities and the NS interior}
\label{Pulsar Timing Irregularities and the NS interior}
The internal structure of a NS is very complicated, with uncertainties in its internal composition. Fig.~\ref{fig1} shows an artist's impression of a pulsar and the zoomed-in structure of its interior. The outer crust of a NS consists of nuclei arranged in a lattice in the background of electrons with densities varying from $\sim10^4$ to $\sim4$ $\times$ $10^{11}$ $g/cm^3$. Neutrons drip out of nuclei when the density reaches $\sim 4$ $\times$ $10^{11}$ $g/cm^3$ and this is the beginning of the inner crust that is composed of neutron-rich nuclei and free neutrons in the background of relativistic electrons \citep{Bandyopadhyay2017}. Free neutrons form Cooper pairs, and the onset of neutron superfluidity begins in the inner crust. The neutron matter in the core also becomes superfluid. Being cold objects, neutron stars support neutron superfluidity in their interiors \citep{alparpines1985}. The composition of the inner core is not properly known since the density of the inner core is much higher than nuclear density. 

A beam of radio signal is emitted from the magnetic pole of the neutron star. The pulsed signal as received by the telescope is very weak and unstable. However, when averaged over several hundred periods \citep{rathnashree1995}, this profile becomes stable and unique to a particular pulsar. This stable profile is known as an integrated pulse profile (shown in the right top corner of Fig.~\ref{fig1}). The rotation of the star can be characterized by associating a time-of-arrival with the fiducial point on this average profile and comparing these observed arrival times to those predicted by a rotational model of the star, using a technique called pulsar timing. 

\begin{figure}[!t]
\includegraphics[width=.99\columnwidth]{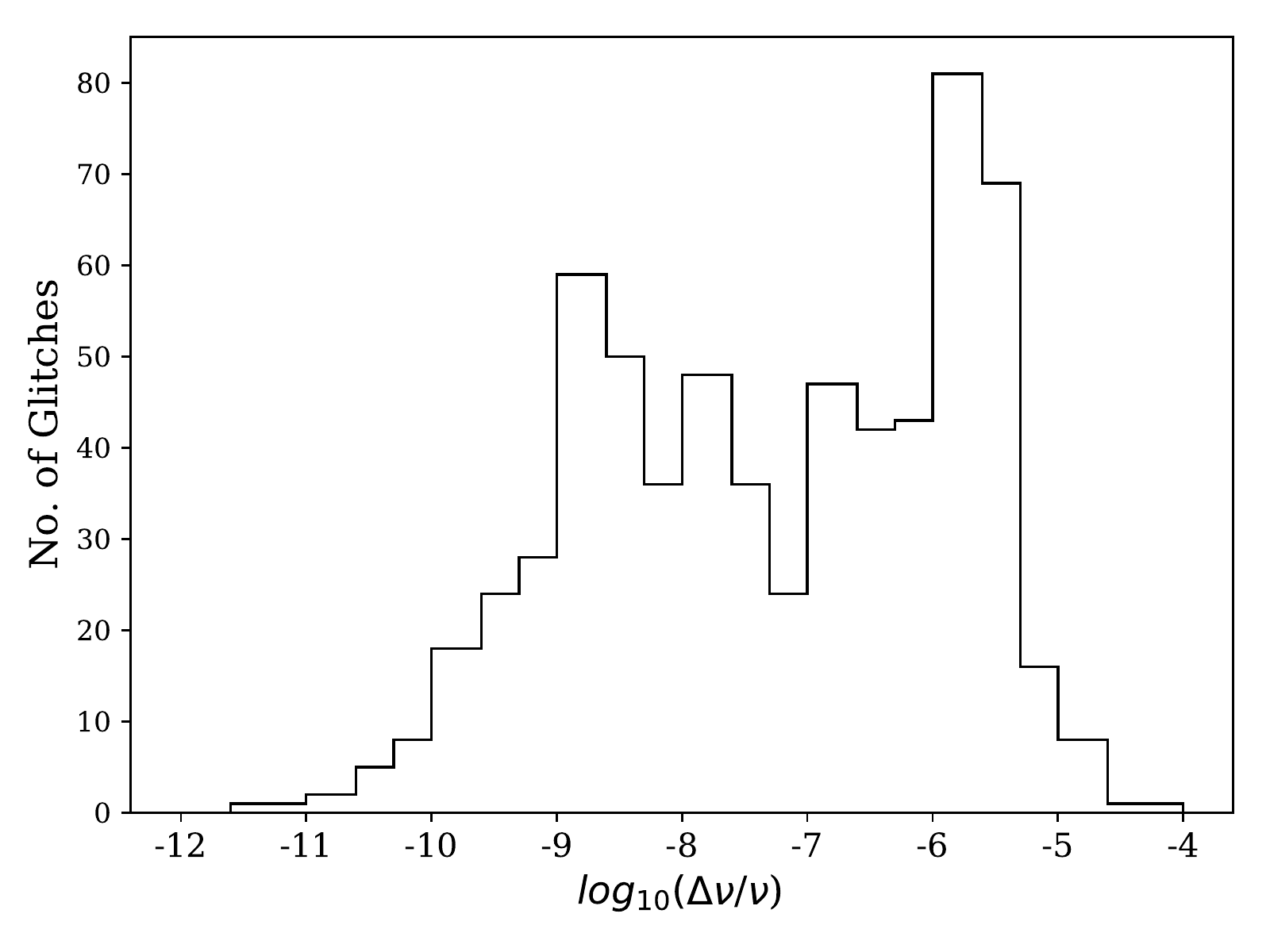}
\caption{The distribution of $log_{10}(\frac{\Delta \nu}{\nu})$ is shown above. The data for this has been taken from Jodrell bank observatory glitch catalogue$^1$ \citep{espinoza2011}. The plot includes 651 glitches in 207 pulsars.}\label{fig2}
\end{figure}

Glitches and timing noise are rotational irregularities seen in pulsars with the help of pulsar timing technique. A glitch is an abrupt increase in the rotation rate of the star, whereas timing noise refers to a random variation in the rotation rate of the star. These phenomena are prominent in young pulsars.  It is believed that the occurrence of pulsar glitches are governed by the internal structure and composition of the NS.

\begin{figure}[!t]
\includegraphics[width=.98\columnwidth]{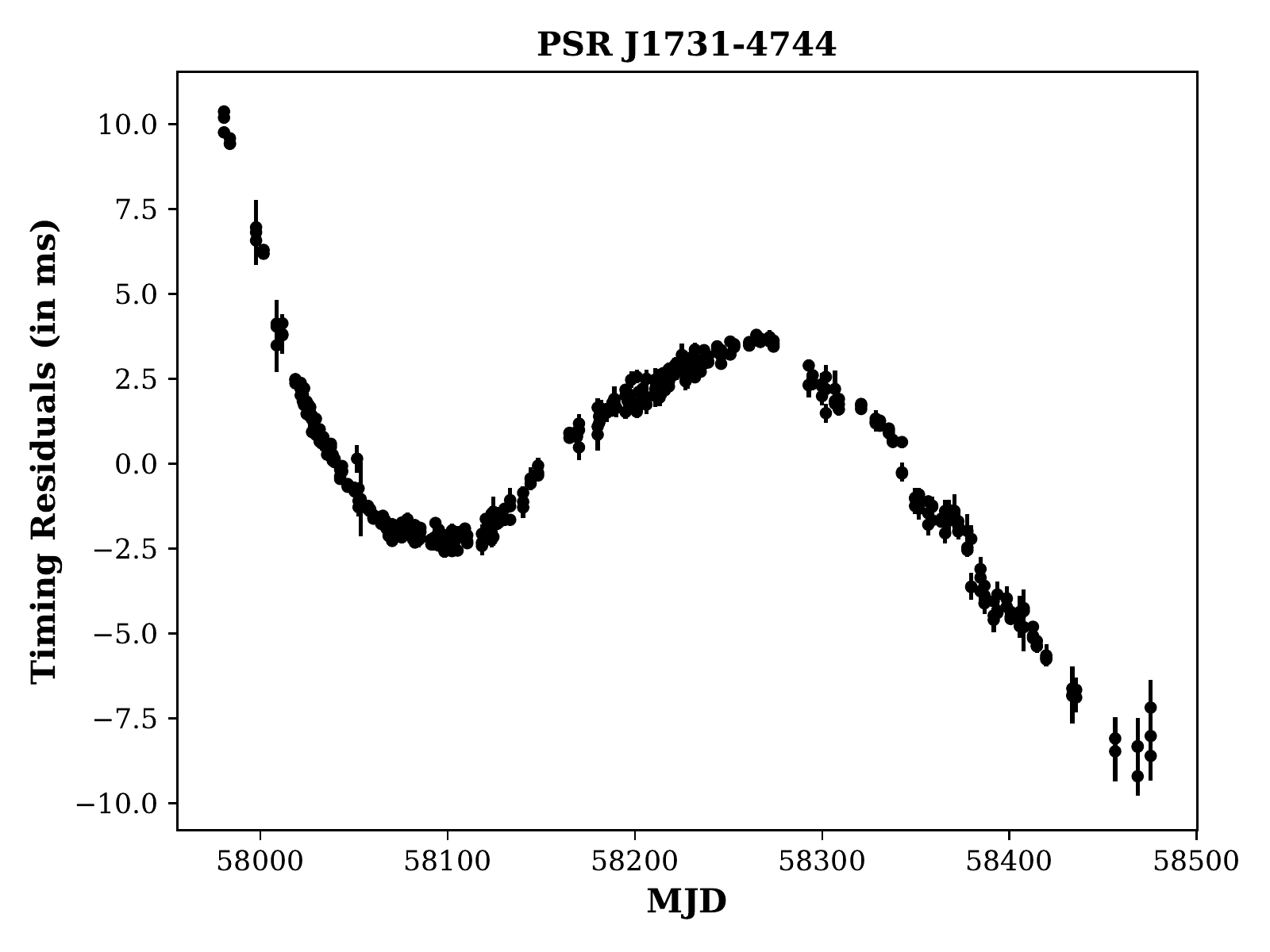}
\caption{The timing residuals seen in PSR 1731-4744 observed with the ORT. It can be seen that the residuals show a trend, which is due to the strong timing noise in this pulsar.}
\label{fig3}
\end{figure}

The initial model of a pulsar glitch considered a quake like event, which was referred to as ``\textit{starquake model}'' \citep{starquakeruderman1969}. In this model, the spin-up was explained by considering the change in the moment of inertia of the star by the sudden relaxation of the neutron star crust from its stressed oblate structure. This model predicted such a starquake to occur after 300 years \citep{RNM+sympo+2018}, whereas the second Vela glitch was observed after 2.5 years by \cite{Reichley+Downs+1971}, which ruled out the starquake mechanism to explain the Vela glitches. The most accepted model today relies on superfluid vortices inside the neutron star crust and their dynamics \citep{AndersonItoh1975,AlparvortexcreepII1984,AlparvortexcreepI1984}. With temperatures typically of the order $10^7K$ inside the NS \citep{Sauls1989}, the highly degenerate matter (mostly neutrons) will be in a superfluid state. Since superflow is a potential flow, it cannot support circulation (rotation), so it breaks up into quanta of vortices. It is energetically favourable for the vortices to pin to nuclear clusters at the lattice sites. The superfluid angular velocity becomes constant due to such a fixed number of the vortices. The crust continuously slows down due to electromagnetic torques, but the superfluid component maintains its angular momentum resulting in a rotation lag between the two. When this lag becomes equal to a critical lag, the Magnus forces unpins the vortices, and a large amount of angular momentum gets transferred to the crust catastrophically and spins it up. This disturbance in the rotation of the crust is reflected in the arrival times of the individual pulses, indicating a glitch.  

An interesting feature of pulsar glitches is the bimodality of glitch sizes $(\frac{\Delta \nu}{\nu})$ \citep{Eya_2019}, shown in Fig.~\ref{fig2}. $\nu$ is the rotational frequency of the pulsar and $\Delta \nu$ is the change in the rotational frequency during a glitch and is defined as $(\nu_{post-glitch} - \nu_{pre-glitch})$. The data for this plot has been taken from the Jodrell Bank observatory glitch catalogue\footnote{http://www.jb.man.ac.uk/pulsar/glitches/gTable.html} \citep{espinoza2011} and 651 glitches in 207 pulsars have been used. The bimodality indicates that the nature of glitches may be of two types: large and small. Although the cause of this dichotomy is still unknown, this may be because of the origin of large and small glitches from different parts of the NS interior. However, it is not clearly known if there should be two different models for these two types of glitches.

The migration and repinning of the vortices defines the post-glitch behaviour. The post-glitch relaxation phase is characterized by an exponential or linear recovery or a combination of the two,  often with single or multiple components \citep{Yu2013}. The timescale of relaxation helps to probe the coupling between the normal component and superfluid component in the inner crust of the star. The relaxation period could last for days or even months. The coupling strength, in turn, is temperature and pinning force-dependent quantity. In between the glitches, there occurs a spin-down of the superfluid in the vortex creep method. In this model, there is an average flow of vortices due to thermal activation against the pinning barrier. Due to the sudden changes in rotation rates during a glitch, the vortex creep gets affected. The response of the vortex creep to the glitch induced changes can be studied with the help of post-glitch relaxation \citep{AlparvortexcreepI1984}. There can be other effects like entrainment \citep{Nchamel2013}, a non-dissipative coupling of interpenetrating superfluid neutrons and the electron-proton normal fluid, which can bring other intricacies to the problem. 

\begin{figure*}[ht]
\centering
\includegraphics[scale=0.75]{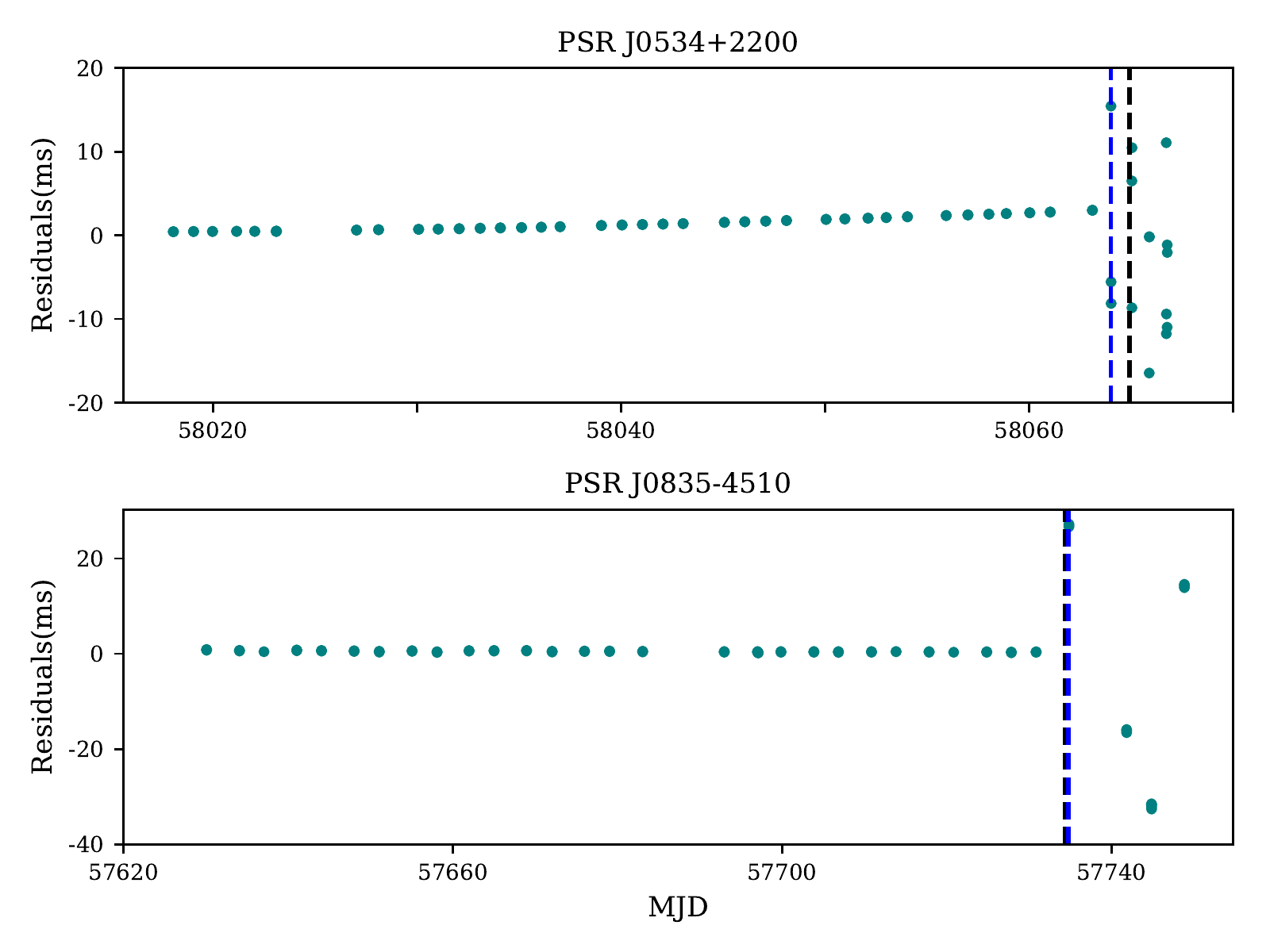}
\caption{This figure shows the timing residuals with glitches used for testing the Automated Glitch Detection Pipeline (AGDP) implemented at the Ooty Radio Telescope. The blue dotted lines depict the epoch where AGDP gives an alarm of probable glitch detection. The scatter in the points after the glitch is due to the lack of phase connection since the detection is in real time. The black dotted lines represent the glitch epochs as given in \cite{basu2020}. The details of such tests are provided in \cite{agdp}.}
\label{fig4}
\end{figure*}

Glitches will help us better understand the interior of neutron stars. In this regard, a quantity known as the fractional moment of inertia (FMI), which is the ratio of the moment of inertia of the crust to the total moment of inertia is very important. The FMI can also be estimated theoretically for a different equation of state (EoS) and hence glitches are also indirect probes to the NS EoS. It has been seen that the FMI estimated theoretically for different EoSs, do not agree with the observed FMI \citep{Basu_2018, singha2022}. It has also been suggested that the superfluidity inside the neutron star core needs to be invoked to explain a few of the observed large glitches \citep{Montoli+2020, Pizzochero+2020, Basu_2018}.

A subclass of pulsar glitches is a slow glitch \citep{slowgl1, slowgl2, Zhou2019}, which is characterised with a continuous increase in frequency over several hundreds of days as also seen in Fig. \ref{fig6}. Slow glitches are very rare, and a total of around 30 slow glitches have been detected so far \citep{Zhou2019}. There have been several attempts to model and explain these unusual kinds of glitches \citep{slowgl3, slowgl4} both theoretically and phenomenologically. A more detailed analysis along with advancements in observations will help us probe the glitch rise phenomena better, which will therefore help in studying the neutron star microphysics \citep{Graber_2018}.

Timing noise can be identified as a red-noise process in the time of arrival of the pulses due to an auto-correlated process, which happens on a time scale of months to years. Fig.~\ref{fig3} shows the timing residuals of PSR J1731$-$4744 with high timing noise observed using the Ooty Radio Telescope. The first attempt to describe timing noise was done by \cite{Boynton_1972} for PSR J0534+2200 (Crab pulsar) attributing it to random walks in any of the following: phase, frequency or the pulsar spin-down rate. With decades of studies, it has been understood that the timing noise is due to different random walks in spin, spin-derivatives, and jumps in phase and spin parameters. However, the theoretical origin of the timing noise is still not properly known. \cite{Jones1990} suggest that timing noise originates due to the modulation in the coupling of the neutron star crust and the superfluid core. \cite{lyne2010} showed that in many pulsars, the timing noise is correlated to the magnetospheric fluctuations. There have been other proposed theories too, which we do not cover in this short review. Considering timing noise to be a red noise process, its power spectrum density can be written as ~\citep{parthasarathy2019},
\begin{equation}
\centering
\label{tneqn}
\mathcal{P}(f) = A\, \frac{(f_c/f_{yr})^{-\beta} }{\Big[1 + (f/f_c)^{-\beta/2}\Big]^2}
\end{equation}
Here, $A = A_{red}^2/12\pi^2$ and $A_{red}$ being the red-noise amplitude (RedAmp), $\beta$ is the index (RedSlope) and $f_c$ is the corner frequency in the units of year $^{-1}$.

\section{Glitch Monitoring Programs in India}
\label{Glitch Monitoring Programs in India}
There are two large radio telescopes in India: the Ooty Radio Telescope  \citep[ORT : ][]{swarup1971large} and the upgraded Giant Metrewave Radio Telescope  \citep[uGMRT : ][]{ugmrt}, both of which have produced very good results over the years. These telescopes have also been used to monitor a sample of glitching pulsars in order to detect and study glitches and timing noise.

The ORT is a 530 m long and 30 m wide parabolic dish located on a north-south hill of 11$^0$ latitude. Its observing frequency is 326.0 MHz. Using the ORT, monitoring of bright sources like J0534+2200 (Crab pulsar) and J0835-4510 (Vela pulsar) began since 2014. This was followed by a high cadence monitoring of a sample of 11 frequently glitching pulsars since 2016. Recently, the sample has been increased to 14. Since the telescope operates at low frequency, we selected pulsars with low Dispersion Measure (DM) to avoid excessive scatter-broadened pulse profiles. These observations at the ORT have a cadence of around 1-7 days. The ORT pulsar back end PONDER \citep{Naidu_2015} converts the raw data into time-series data. De-dispersion is done with respect to the highest frequency of the band, i.e. 334.5 MHz. A real-time automated glitch detection pipeline (AGDP) \citep{agdp} has been developed and implemented at ORT. AGDP generate timing residuals using the time-series data from the ORT with the help of various pulsar softwares like DSPSR \citep{DSPSR}, PSRCHIVE \citep{PSRCHIVE}, TEMPO2 \citep{tempo2I, Tempo2II}, etc. Based on a pre-defined criteria, AGDP tries to find if there is a possibility of a   glitch in the most recently observed epoch using these residuals. It implements a polynomial fitting technique to detect glitches in real-time and send alarms to the observer. Fig.~\ref{fig4} shows the implementation of AGDP on real glitches. In the figure, the blue dotted lines depict the epoch where the pipeline gives an alarm for the probable detection of glitches in real-time, and the black ones represent the real glitch epoch given in \cite{basu2020}. A detailed description of these tests are presented in \cite{agdp}. AGDP will help us allow control the cadence of observations and also study the glitch recoveries better. 

The uGMRT is an interferometer with a wide range of frequency coverage. It consists of 30 antennas (each of 45 m diameter) spanning over 25 km. There are four observing bands spanning from 250 MHz to 1450 MHz. It can be used to monitor a large sample of glitching pulsars in a modest time to obtain enough signal to noise (S/N) ratio on pulse profile to perform the timing studies. A sample of high glitch rate pulsars has been selected, and a successful monitoring program has been running using uGMRT since 2016. These observations are done at a typical cadence of around 10-14 days. The raw data collected from uGMRT is reduced to PSRFITS files using the \texttt{pinta} pipeline~\citep{pinta}, which has been initially developed for the Indian Pulsar Timing Array (InPTA) experiment \citep{JoshiAB+18}. The PSRFITS files generated are used to do further analysis using PSRCHIVE and TEMPO2 in order to generate timing residuals. AGDP is run on these residuals offline.

\begin{figure}[!t]
\includegraphics[width=.95\columnwidth]{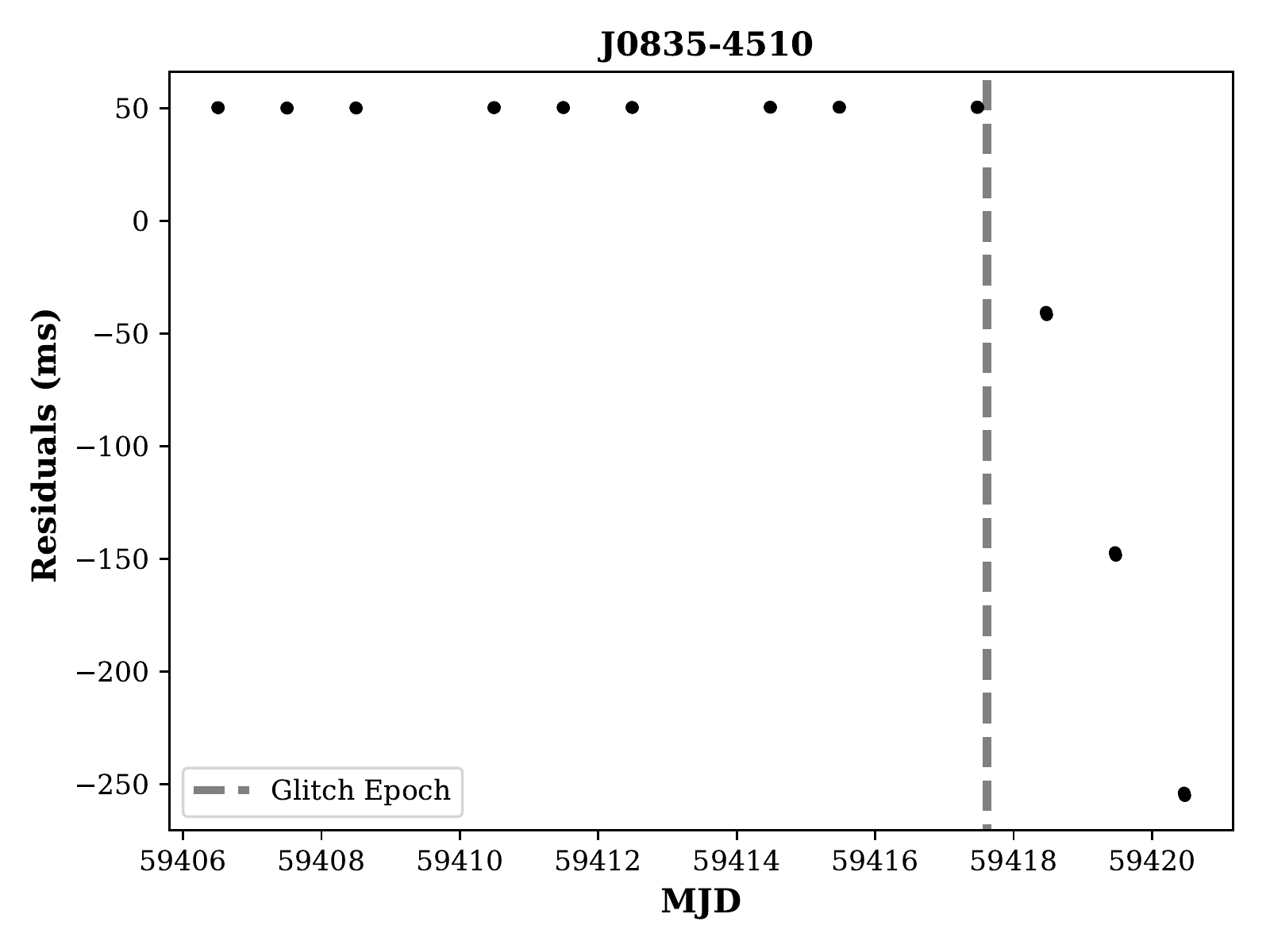}
\caption{The glitch seen in PSR J0835$-$4510 (Vela pulsar) observed with the ORT as presented in \cite{AtelVela}}\label{fig5}
\end{figure}

\begin{figure}
\centering
\includegraphics[width=.98\columnwidth]{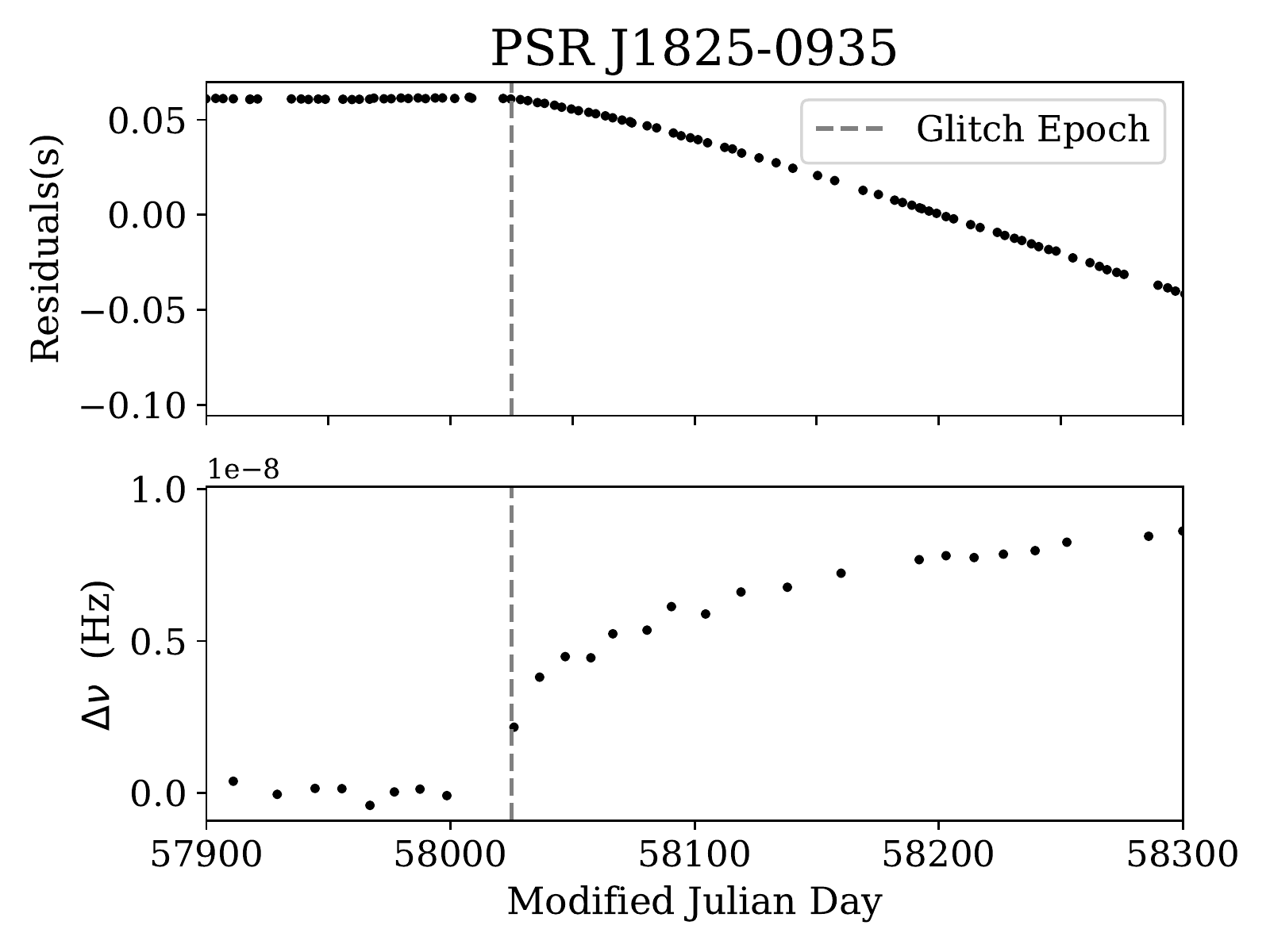}
\caption{A slow glitch seen in PSR J1825$-$0935. The upper plot shows the timing residuals and the lower plot represents the frequency evolution during the glitch.}\label{fig6}
\end{figure}

\begin{figure*}
\centering
\includegraphics[scale=0.6]{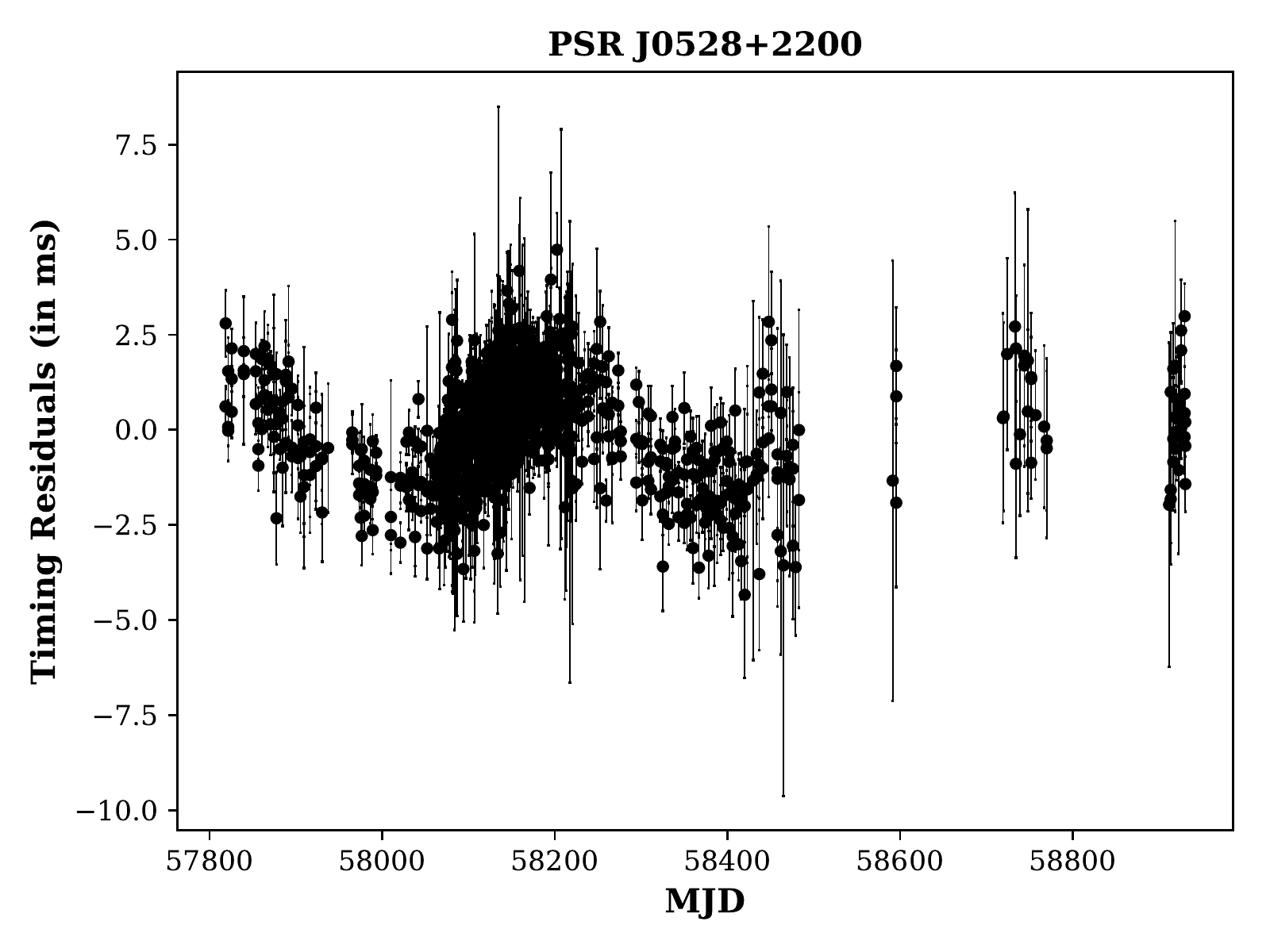}\\
\includegraphics[scale=0.6]{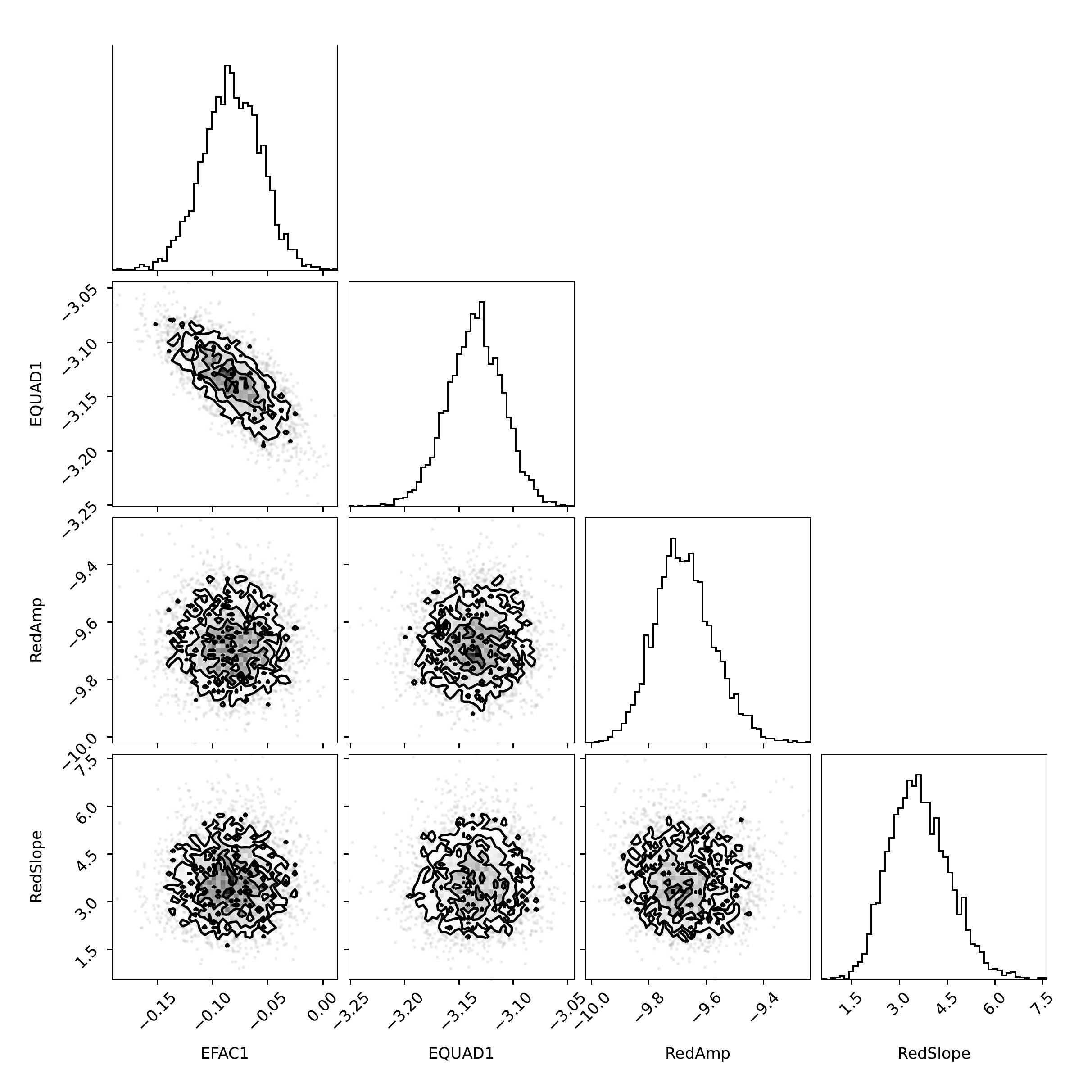}
\caption{Timing Noise analysis of PSR J0528+2200. The upper plot shows the timing residuals of this pulsar observed with the ORT. The lower plot represents the posteriors obtained using \texttt{TempoNest} \citep{Lentati_2013}. \texttt{TempoNest} is a Bayesian analysis software which uses MultiNest \citep{feroz2009,cameron2013} to perform nested sampling \citep{nested} and obtain the timing noise parameters. Here RedAmp and RedSlope are the red noise amplitude and index as given in eqn.~\ref{tneqn}. EFAC1 and EQUAD1 are the white noise parameters. Further details of such analysis will be provided in Singha et al. (in preparation).}
\label{fig7}
\end{figure*}

The glitch monitoring programs using the uGMRT and ORT have already produced good results. The first results of this monitoring program were reported in \cite{basu2020} presenting 11 glitches in 8 young pulsars. Out of these, 3 of the glitches had not been reported elsewhere. A detailed modelling of the rise time of the largest glitch seen in the PSR J0534+2200 (Crab pulsar) was also presented in this paper. It is important to note here that even a very small glitch of $\sim 10^{-9}$ was also detected in this program. A second paper (Singha et al., in preparation) with results of timing irregularities will soon be submitted. This will include around 10 new detected glitches and also timing noise studies of around 11 pulsars with the same collected timing data. A detailed analysis of the recent glitch seen in PSR J0835-4510 \citep[Vela pulsar : ][]{AtelVela} (shown in Fig.~\ref{fig5}) will also be presented. We have also detected a slow glitch in our monitoring program. Fig. \ref{fig6} shows the timing residuals of a slow glitch detected in J1825$-$0935. Fig \ref{fig7} shows the preliminary timing noise analysis for pulsar J0528+2200 observed using the ORT. The upper plot is the timing residuals of the pulsar and the lower plot represents the posterior plot obtained for the red noise parameters. The timing noise analysis has been done using the software, \texttt{TempoNest} \citep{Lentati_2013}. It utilises nested sampling \citep{nested} with the help of MultiNest \citep{feroz2009,cameron2013} to find the parameters. 

On the theoretical front too, the Indian NS community has made significant contributions in understanding the origin of the timing irregularities in NSs. With the help of FMI estimated for different equations of states of NSs, \cite{Basu_2018} and \cite{singha2022} showed that all the glitches observed cannot be explained by the transfer of angular momentum from the inner crust. \cite{singha2022} also demonstrated this with rotating NSs. Further studies are being planned in order to understand the crust and core contributions to glitches better. There are plans to study the transient phenomena like the glitch rise time in the framework of two-fluid models of NSs. A full theoretical approach to timing noise has not been achieved yet. However, modelling timing noise as a red noise process with the help of Bayesian analysis will also give an insight into the complex behaviour of young pulsars. Along with such approaches, studying timing noise in some phenomenological models are also planned.

\section{Pulsar Observations with the SKA}
\label{Pulsar Observations}
The SKA project\footnote{https://www.skatelescope.org/the-ska-project/} aims to build the world’s largest radio telescope, with a square kilometre of collecting area and a maximum baseline of 3000+ km. It will cover the frequency range of 50 MHz to 20 GHz. It will be about 50-100 times more powerful for spectral line observations and 1000 times more powerful for continuum observations than present radio telescopes \citep{Fcombes2015}. The SKA telescope consists of two instruments, the SKA-low and the SKA-mid, which are currently under construction in Australia and South Africa, respectively. Phase I of the SKA telescope will provide approximately 10 percent of the collecting area of the full telescope and is briefly described below.

In South Africa, the SKA’s mid-frequency telescope will consist of 133 15-m dishes with the existing 64 13.5m MeerKAT \citep{meerkat} dishes. The majority of the dishes will be in a central core region, with a few in three spiral arms extending over 150 km. The SKA-mid telescope will cover a frequency range from 0.35 GHz to around 20 GHz, whereas 512 stations, each containing 256 individual antennas, will be arranged in a large core with three spiral arms scattered over a distance of 65 km to form the SKA-low telescope in Australia. It will cover a frequency range of 50$-$350 MHz.  

Pulsar astronomy mostly depends on radio telescopes. The SKA will play a crucial role in pulsar timing studies, including those focusing on the timing irregularities in pulsars. As the SKA telescope is an interferometer, it is a flexible instrument for such studies. Apart from its large collecting area, the beam-former of the SKA telescope provides 16 concurrent beams, which can be formed by different combinations of SKA antennas/stations, called sub-arrays, for both the SKA-low and the SKA-mid. The observations of brighter pulsars, such as PSR J0534+2200 (Crab pulsar) and J0835-4510 (Vela pulsar), can be done with even a single antenna. This allows a dense observing campaign of these frequently glitching pulsars, which exhibit small and large glitches, respectively. The fainter pulsars can be observed with sub-arrays consisting of an increasing number of antennas. Multiple young pulsars can be observed concurrently, optimising the total observing time required for a glitch monitoring campaign for a large sample of pulsars. Lastly, the observing strategy can also be tuned by employing different beams at different frequency bands available with nearby and fainter pulsars being observed at a lower frequency and the more distant pulsars being observed at higher frequencies.

One of the major goals of the SKA is to search for new pulsars in the galaxy. The SKA is expected to add around 14000 normal pulsars and 6000-millisecond pulsars \citep{smits} to the currently known population of around 3300 pulsars \citep[ATNF pulsar catalogue : ][]{Manchester_2005}\footnote{https://www.atnf.csiro.au/research/pulsar/psrcat/}. The SKA will, therefore, significantly increase the sample of glitching pulsars from the currently reported sample of 207 pulsars showing 651 glitches\footnote{http://www.jb.man.ac.uk/pulsar/glitches/gTable.html}. The SKA will significantly improve both of these counts, resulting in a collection of extensive observational data of glitches and post-glitch recovery.

The beamformer and central signal processing of the SKA-telescope will provide time-stamped integrated pulse profile (IPP) in real-time. Implementation of real-time glitch detection techniques in the science data processor element of the SKA-telescope has the potential for providing real-time glitch triggers, much in the same way as transient events will produce triggers for the transient science with the SKA telescope. This opens the possibility of high cadence post-glitch recovery observations of a glitching pulsar in a target of opportunity mode. Such monitoring has been very limited in the past and will be critical in understanding the underlying physics of a glitch recovery.

Some of the major goals of a program to monitor timing irregularities in young pulsars with the SKA telescope are:
\begin{enumerate}
    \item Detection of glitches in a large sample of pulsars. 
    \item Study of post-glitch behaviour.
    \item Characterization of timing noise.
    \item Catching and investigating rare events, such as post-glitch rise time and slow glitches.
\end{enumerate}
The appropriate observing strategies to achieve these goals can be based on observations from the existing SKA path-finder telescope. In the next section, these are outlined based on the experience derived from a similar program with the uGMRT.

\section{Lessons from current pathfinder programs}
\label{Lessons}
The uGMRT \citep{ugmrt} due to its seamless coverage of a broad frequency range, large bandwidth, multiple frequencies of operation and high sensitivity phased array capabilities have been given the status of SKA-pathfinder. The telescope can be used as multiple telescope sub-arrays as it is an interferometer. Thus, the uGMRT has notable similarities with the SKA.

Glitch monitoring programs using the uGMRT and ORT have been successfully running for the past few years. These programs have not just contributed to the overall glitch statistics, but they have also helped in testing observing strategies and developing novel techniques to detect glitches in real-time as described in Section \ref{Glitch Monitoring Programs in India} These programs utilize multiple bands of the uGMRT and the ORT to optimise the observing time on one hand and improved cadence on the other hand. Our sample of more than 20 glitching pulsars with a glitch detection rate of 0.38 glitches per year per pulsar was divided into three groups. Pulsars with DM less than about 30 pc\,cm$^{-3}$ 
were observed at 326.0 MHz, whereas fainter pulsars and pulsars with DM between 40 to 150 pc\,cm$^{-3}$ were observed with Band 4 ( 550$-$850 MHz) of the uGMRT. All pulsars with DM greater than 150  pc\,cm$^{-3}$ were observed with 400 MHz band-pass at Band 5(1060$-$1460 MHz) at the uGMRT. This strategy helps in avoiding the reduction of detected signal-to-noise ratio due to the expected pulse scatter-broadening in high DM pulsar. It may be noted that larger scatter-broadening also increases timing uncertainty. Moreover, the steep spectrum of pulsars implies that a short integration time is needed even for faint pulsars at lower frequencies. Thus, one requires much smaller observing time compared to a conventional monitoring program. This allowed us to carry out two observations per month (for pulsars observed using the uGMRT) and a better cadence than the conventional monitoring program. The observing time can be reduced even further by employing multiple sub-arrays for multiple targets using 2 to 3 beams of the uGMRT. While this strategy has not been tried by us, we plan to test this in upcoming observations. With 16 beams available in the SKA-mid and the SKA-low telescopes and multiple combinations of antennas possible for 16 sub-arrays, an optimised program based on our experience can cover a much larger sample of pulsars with the SKA. Work is in progress to evaluate these strategies based on the experience gained from the uGMRT.

An offline analysis pipeline was developed for a precision timing program with the GMRT \citep{pinta} to rapidly analyse the data in the uGMRT program. The SKA will use a real-time version of a similar pipeline with data products identical to our pipeline. \cite{agdp} developed a  post-processing pipeline for real-time automated glitch detection (AGDP) for the ORT. This pipeline runs in real-time at the ORT and generates an email trigger for every potential glitch. A rapid target of opportunity request is generated at the uGMRT in the event of a trigger for increased cadence observing of the post-glitch behaviour of the pulsar. This pipeline is compatible with the output product of the central signal processing element of the SKA. We believe it can be easily integrated in the science data processor element of the SKA and generate triggers similar to the transient triggers. The SKA telescope manager can be programmed to initiate follow up high cadence observations in the event of a valid trigger with pre-programmed sub-array and band configurations. This will be very valuable in catching short time-scale recoveries and provide critical data to constrain the recovery physics.

A key factor in determining a valid glitch in AGDP is the amplitude of the timing noise in the pulsar being monitored. While this is a required input for the pipeline, intelligence can be built in the pipeline based on Bayesian analysis of timing residuals for each pulsar during intervals free of glitches. These estimates can be updated in real-time after each observation by incorporating this analysis in the science data processor. While this will improve the effectiveness and precision of AGDP on one hand, it will provide timing noise models of all these young pulsars as a side product. Such trials are currently ongoing in our program and can be very useful for the SKA.

Last, but not the least, these monitoring programs with the uGMRT and ORT have been useful in developing the SKA user community. They have connected astronomers and astrophysicists in the Indian NS community with both observational and theoretical expertise through the uGMRT and ORT programs. They have launched a theoretical exploration of the equation of state and internal structure of NS by the community and motivated the community to develop interpretation for data in the SKA era. Three Indian students were trained as doctoral students in this field, with further expansion possible in future. Together these efforts would expand the SKA user community with additions of NS astronomers in the area of irregularities in pulsar timing.
It is important to note that an ongoing study using MeerKAT called the thousand pulsar array program \citep{meerkat_pt} aims to observe 1000 pulsars to obtain high precision pulse profiles, half of the pulsars over multiple epochs, and a sequence of single pulse trains. Study of glitches in these pulsars can be a probable project.

\section{Conclusions}
\label{Conclusion}

The SKA will give birth to a new era in radio astronomy. With the help of this mega telescope, the neutron star community will be greatly benefited in particular. Capacity building in India with glitch monitoring programs have been going on in the last 5 years, where more than 20 glitches were discovered, including the largest glitch in PSR J0534+2200 (Crab pulsar) with a peculiar rise time as well as a slow rise time glitch. In addition, timing noise parameters for 11 young pulsars have been determined. Optimal observing strategies using multiple frequency bands and sub-arrays were developed for the uGMRT apart from an automated glitch detection pipeline for the ORT. Lastly, theoretical calculations for different equations of state for both static and rotating NS were carried out to interpret glitch data, and future work is planned for understanding the physics of glitch recovery.

As the construction of the Phase I telescope proceeds, a similar program needs to be designed for the pulsar key science project. The required observations strategies for such a program can be informed for the uGMRT program. This is an optimisation program over different elements of the SKA-baseline design, and work is in progress for such calculations. The AGDP can provide a template for triggered high cadence post-glitch observations with the SKA as described in this paper, but this can be made more effective with simultaneous update of timing noise models. A work in this direction is also planned. Last but not least, the trained person-power available from the Indian program promises to help in effective scientific utilisation of the SKA telescope when it is ready and operational in about 4 years.

\section*{Acknowledgements}

The authors acknowledge the help and support provided by the staff at Radio Astronomy Centre, Ooty and the upgraded Giant Meterwave Radio Telescope during these observations. The ORT and the uGMRT are operated by the National Centre for Radio Astrophysics.

\vspace{-1em}



\bibliography{references}

\begin{thebibliography}{}
\expandafter\ifx\csname natexlab\endcsname\relax\def\natexlab#1{#1}\fi

\bibitem[{{Alpar} {$et~al$.}(1984{\natexlab{a}}){Alpar}, {Anderson}, {Pines},
  \& {Shaham}}]{AlparvortexcreepII1984}
{Alpar}, M.~A., {Anderson}, P.~W., {Pines}, D., \& {Shaham}, J.
  1984{\natexlab{a}}, \apj, 278, 791

\bibitem[{{Alpar} {$et~al$.}(1984{\natexlab{b}}){Alpar}, {Pines}, {Anderson},
  \& {Shaham}}]{AlparvortexcreepI1984}
{Alpar}, M.~A., {Pines}, D., {Anderson}, P.~W., \& {Shaham}, J.
  1984{\natexlab{b}}, \apj, 276, 325

\bibitem[{{Anderson} \& {Itoh}(1975)}]{AndersonItoh1975}
{Anderson}, P.~W., \& {Itoh}, N. 1975, Nature, 256, 25

\bibitem[{{Baade} \& {Zwicky}(1934)}]{Baade}
{Baade}, W., \& {Zwicky}, F. 1934, Physical Review, 46, 76

\bibitem[{Bailes {$et~al$.}(2018)Bailes, Barr, Bhat, Brink, Buchner, Burgay,
  Camilo, Champion, Hessels, Janssen, Jameson, Johnston, Karastergiou,
  Karuppusamy, Kaspi, Keith, Kramer, McLaughlin, Moodley, Oslowski, Possenti,
  Ransom, Rasio, Sievers, Serylak, Stappers, Stairs, Theureau, van Straten,
  Weltevrede, \& Wex}]{meerkat_pt}
Bailes, M., Barr, E., Bhat, N. D.~R., {$et~al$.} 2018,
  doi:10.48550/ARXIV.1803.07424

\bibitem[{Bandyopadhyay(2017)}]{Bandyopadhyay2017}
Bandyopadhyay, D. 2017, Journal of Astrophysics and Astronomy, 38, 37

\bibitem[{Basu {$et~al$.}(2018)Basu, Char, Nandi, Joshi, \&
  Bandyopadhyay}]{Basu_2018}
Basu, A., Char, P., Nandi, R., Joshi, B.~C., \& Bandyopadhyay, D. 2018, The
  Astrophysical Journal, 866, 94

\bibitem[{Basu {$et~al$.}(2019)Basu, Joshi, Krishnakumar, Bhattacharya, Nandi,
  Bandhopadhay, Char, \& Manoharan}]{basu2020}
Basu, A., Joshi, B.~C., Krishnakumar, M.~A., {$et~al$.} 2019, Monthly Notices
  of the Royal Astronomical Society, 491, 3182

\bibitem[{{Boynton} {$et~al$.}(1972){Boynton}, {Groth}, {Hutchinson}, {Nanos},
  {Partridge}, \& {Wilkinson}}]{Boynton_1972}
{Boynton}, P.~E., {Groth}, E.~J., {Hutchinson}, D.~P., {$et~al$.} 1972, ApJ,
  175, 217

\bibitem[{Cameron \& Pettitt(2014)}]{cameron2013}
Cameron, E., \& Pettitt, A. 2014, Statistical Science, 29, 397

\bibitem[{Chamel(2013)}]{Nchamel2013}
Chamel, N. 2013, Phys. Rev. Lett., 110, 011101

\bibitem[{Combes(2015)}]{Fcombes2015}
Combes, F. 2015, Journal of Instrumentation, 10, C09001–C09001

\bibitem[{{Cordes} \& {Helfand}(1980)}]{Cordes_Helfand1980}
{Cordes}, J.~M., \& {Helfand}, D.~J. 1980, ApJ, 239, 640

\bibitem[{{Edwards} {$et~al$.}(2006){Edwards}, {Hobbs}, \&
  {Manchester}}]{Tempo2II}
{Edwards}, R.~T., {Hobbs}, G.~B., \& {Manchester}, R.~N. 2006, MNRAS, 372, 1549

\bibitem[{{Epstein} \& {Baym}(1988)}]{BaymEipstein1988}
{Epstein}, R.~I., \& {Baym}, G. 1988, ApJ, 328, 680

\bibitem[{{Espinoza} {$et~al$.}(2011){Espinoza}, {Lyne}, {Stappers}, \&
  {Kramer}}]{espinoza2011}
{Espinoza}, C.~M., {Lyne}, A.~G., {Stappers}, B.~W., \& {Kramer}, M. 2011,
  MNRAS, 414, 1679

\bibitem[{Eya {$et~al$.}(2019)Eya, Urama, \& Chukwude}]{Eya_2019}
Eya, I.~O., Urama, J.~O., \& Chukwude, A.~E. 2019, Research in Astronomy and
  Astrophysics, 19, 089

\bibitem[{Feroz {$et~al$.}(2009)Feroz, Hobson, \& Bridges}]{feroz2009}
Feroz, F., Hobson, M.~P., \& Bridges, M. 2009, Monthly Notices of the Royal
  Astronomical Society, 398, 1601

\bibitem[{Graber {$et~al$.}(2018)Graber, Cumming, \& Andersson}]{Graber_2018}
Graber, V., Cumming, A., \& Andersson, N. 2018, The Astrophysical Journal, 865,
  23

\bibitem[{{Gupta} {$et~al$.}(2017){Gupta}, {Ajithkumar}, {Kale}, {Nayak},
  {Sabhapathy}, {Sureshkumar}, {Swami}, {Chengalur}, {Ghosh},
  {Ishwara-Chandra}, {Joshi}, {Kanekar}, {Lal}, \& {Roy}}]{ugmrt}
{Gupta}, Y., {Ajithkumar}, B., {Kale}, H.~S., {$et~al$.} 2017, Current Science,
  113, 707

\bibitem[{Gügercinoğlu {$et~al$.}(2022)Gügercinoğlu, Ge, Yuan, \&
  Zhou}]{erbil2022}
Gügercinoğlu, E., Ge, M.~Y., Yuan, J.~P., \& Zhou, S.~Q. 2022, Monthly
  Notices of the Royal Astronomical Society, 511, 425

\bibitem[{{Haskell} {$et~al$.}(2018){Haskell}, {Khomenko}, {Antonelli}, \&
  {Antonopoulou}}]{haskell+2018}
{Haskell}, B., {Khomenko}, V., {Antonelli}, M., \& {Antonopoulou}, D. 2018,
  arXiv e-prints, arXiv:1806.10168

\bibitem[{Hessels {$et~al$.}(2006)Hessels, Ransom, Stairs, Freire, Kaspi, \&
  Camilo}]{1.4mspulsar}
Hessels, J. W.~T., Ransom, S.~M., Stairs, I.~H., {$et~al$.} 2006, Science, 311,
  1901–1904

\bibitem[{Hobbs {$et~al$.}(2006b)Hobbs, Lyne, \& Kramer}]{Hobbs_2006}
Hobbs, G., Lyne, A., \& Kramer, M. 2006b, Chinese Journal of Astronomy and
  Astrophysics, 6, 169

\bibitem[{{Hobbs} {$et~al$.}(2006a){Hobbs}, {Edwards}, \&
  {Manchester}}]{tempo2I}
{Hobbs}, G.~B., {Edwards}, R.~T., \& {Manchester}, R.~N. 2006a, MNRAS, 369, 655

\bibitem[{Jonas(2009)}]{meerkat}
Jonas, J.~L. 2009, Proceedings of the IEEE, 97, 1522

\bibitem[{{Jones}(1990)}]{Jones1990}
{Jones}, P.~B. 1990, \mnras, 246, 364

\bibitem[{{Joshi} {$et~al$.}(2018){Joshi}, {Arumugasamy}, {Bagchi},
  {Bandyopadhyay}, {Basu}, {Dhanda Batra}, {Bethapudi}, {Choudhary}, {De},
  {Dey}, {Gopakumar}, {Gupta}, {Krishnakumar}, {Maan}, {Manoharan}, {Naidu},
  {Nandi}, {Pathak}, {Surnis}, \& {Susobhanan}}]{JoshiAB+18}
{Joshi}, B.~C., {Arumugasamy}, P., {Bagchi}, M., {$et~al$.} 2018, Journal of
  Astrophysics and Astronomy, 39, 51

\bibitem[{Lentati {$et~al$.}(2013)Lentati, Alexander, Hobson, Feroz, van
  Haasteren, Lee, \& Shannon}]{Lentati_2013}
Lentati, L., Alexander, P., Hobson, M.~P., {$et~al$.} 2013, Monthly Notices of
  the Royal Astronomical Society, 437, 3004–3023

\bibitem[{{Lyne}(2013)}]{Lyne+2013}
{Lyne}, A. 2013, in IAU Symposium, Vol. 291, Neutron Stars and Pulsars:
  Challenges and Opportunities after 80 years, ed. J.~{van Leeuwen}, 183--188

\bibitem[{Lyne {$et~al$.}(2010)Lyne, Hobbs, Kramer, Stairs, \&
  Stappers}]{lyne2010}
Lyne, A., Hobbs, G., Kramer, M., Stairs, I., \& Stappers, B. 2010, Science,
  329, 408

\bibitem[{{Manchester}(2018)}]{RNM+sympo+2018}
{Manchester}, R.~N. 2018, arXiv e-prints, arXiv:1801.04332

\bibitem[{Manchester {$et~al$.}(2005)Manchester, Hobbs, Teoh, \&
  Hobbs}]{Manchester_2005}
Manchester, R.~N., Hobbs, G.~B., Teoh, A., \& Hobbs, M. 2005, The Astronomical
  Journal, 129, 1993

\bibitem[{{Montoli} {$et~al$.}(2020){Montoli}, {Antonelli}, {Magistrelli}, \&
  {Pizzochero}}]{Montoli+2020}
{Montoli}, A., {Antonelli}, M., {Magistrelli}, F., \& {Pizzochero}, P. 2020,
  arXiv e-prints, arXiv:2005.01594

\bibitem[{Naidu {$et~al$.}(2015)Naidu, Joshi, Manoharan, \&
  Krishnakumar}]{Naidu_2015}
Naidu, A., Joshi, B.~C., Manoharan, P.~K., \& Krishnakumar, M.~A. 2015,
  Experimental Astronomy, 39, 319–341

\bibitem[{{Parthasarathy} {$et~al$.}(2019){Parthasarathy}, {Shannon},
  {Johnston}, {Lentati}, {Bailes}, {Dai}, {Kerr}, {Manchester}, {Os{\l}owski},
  {Sobey}, {van Straten}, \& {Weltevrede}}]{parthasarathy2019}
{Parthasarathy}, A., {Shannon}, R.~M., {Johnston}, S., {$et~al$.} 2019, \mnras,
  489, 3810

\bibitem[{Peng \& Xu(2008)}]{slowgl4}
Peng, C., \& Xu, R.~X. 2008, Monthly Notices of the Royal Astronomical Society,
  384, 1034

\bibitem[{{Pines} \& {Alpar}(1985)}]{alparpines1985}
{Pines}, D., \& {Alpar}, M.~A. 1985, Nature, 316, 27

\bibitem[{{Pizzochero} {$et~al$.}(2020){Pizzochero}, {Montoli}, \&
  {Antonelli}}]{Pizzochero+2020}
{Pizzochero}, P.~M., {Montoli}, A., \& {Antonelli}, M. 2020, \aap, 636, A101

\bibitem[{{Radhakrishnan} \& {Manchester}(1969)}]{Radman69}
{Radhakrishnan}, V., \& {Manchester}, R.~N. 1969, Nature, 222, 228

\bibitem[{{Rathnasree} \& {Rankin}(1995)}]{rathnashree1995}
{Rathnasree}, N., \& {Rankin}, J.~M. 1995, \apj, 452, 814

\bibitem[{{Reichley} \& {Downs}(1971)}]{Reichley+Downs+1971}
{Reichley}, P.~E., \& {Downs}, G.~S. 1971, Nature Physical Science, 234, 48

\bibitem[{{Ruderman}(1969)}]{starquakeruderman1969}
{Ruderman}, M. 1969, Nature, 223, 597

\bibitem[{Sauls(1989)}]{Sauls1989}
Sauls, J.~A. 1989, Superfluidity in the Interiors of Neutron Stars (Dordrecht:
  Springer Netherlands), 457--490

\bibitem[{{Shabanova}(1998)}]{slowgl2}
{Shabanova}, T.~V. 1998, \aap, 337, 723

\bibitem[{Shabanova(2005)}]{slowgl1}
Shabanova, T.~V. 2005, Monthly Notices of the Royal Astronomical Society, 356,
  1435

\bibitem[{{Shaw} {$et~al$.}(2018){Shaw}, {Lyne}, {Stappers}, {Weltevrede},
  {Bassa}, {Lien}, {Mickaliger}, {Breton}, {Jordan}, {Keith}, \&
  {Krimm}}]{Shaw+2018}
{Shaw}, B., {Lyne}, A.~G., {Stappers}, B.~W., {$et~al$.} 2018, MNRAS, 478, 3832

\bibitem[{Singha {$et~al$.}(2021a)Singha, Basu, Krishnakumar, Joshi, \&
  Arumugam}]{agdp}
Singha, J., Basu, A., Krishnakumar, M.~A., Joshi, B.~C., \& Arumugam, P. 2021a,
  Monthly Notices of the Royal Astronomical Society, 505, 5488

\bibitem[{{Singha} {$et~al$.}(2021b){Singha}, {Joshi}, {Arumugam}, \&
  {Bandyopadhyay}}]{AtelVela}
{Singha}, J., {Joshi}, B.~C., {Arumugam}, P., \& {Bandyopadhyay}, D. 2021b, The
  Astronomer's Telegram, 14812, 1

\bibitem[{Singha {$et~al$.}(2022)Singha, Sampangi~Raman, \& Kumar}]{singha2022}
Singha, J., Sampangi~Raman, M.~V., \& Kumar, A. 2022, Research in Astronomy and
  Astrophysics

\bibitem[{{Skilling}(2004)}]{nested}
{Skilling}, J. 2004, in American Institute of Physics Conference Series, Vol.
  735, Bayesian Inference and Maximum Entropy Methods in Science and
  Engineering: 24th International Workshop on Bayesian Inference and Maximum
  Entropy Methods in Science and Engineering, ed. R.~{Fischer}, R.~{Preuss}, \&
  U.~V. {Toussaint}, 395--405

\bibitem[{Smits {$et~al$.}(2008)Smits, Kramer, Stappers, Lorimer, Cordes, \&
  Faulkner}]{smits}
Smits, R., Kramer, M., Stappers, B., {$et~al$.} 2008, Astronomy \&
  Astrophysics, 493, 1161–1170

\bibitem[{Susobhanan {$et~al$.}(2021)Susobhanan, Maan, Joshi, Prabu, Desai,
  Nobleson, Susarla, Girgaonkar, Dey, Batra, \& et~al.}]{pinta}
Susobhanan, A., Maan, Y., Joshi, B.~C., {$et~al$.} 2021, Publications of the
  Astronomical Society of Australia, 38, e017

\bibitem[{Swarup {$et~al$.}(1971)Swarup, Sarma, Joshi, Kapahi, Bagri, Damle,
  Ananthakrishnan, Balasubramanian, Bhave, \& Sinha}]{swarup1971large}
Swarup, G., Sarma, N., Joshi, M., {$et~al$.} 1971, Nature Physical Science,
  230, 185

\bibitem[{Tan {$et~al$.}(2018)Tan, Bassa, Cooper, Dijkema, Esposito, Hessels,
  Kondratiev, Kramer, Michilli, Sanidas, Shimwell, Stappers, van Leeuwen,
  Cognard, Grie{\ss}meier, Karastergiou, Keane, Sobey, \&
  Weltevrede}]{23.5spulsar}
Tan, C.~M., Bassa, C.~G., Cooper, S., {$et~al$.} 2018, The Astrophysical
  Journal, 866, 54

\bibitem[{{van Straten} \& {Bailes}(2011)}]{DSPSR}
{van Straten}, W., \& {Bailes}, M. 2011, \pasa, 28, 1

\bibitem[{{van Straten} {$et~al$.}(2012){van Straten}, {Demorest}, \&
  {Oslowski}}]{PSRCHIVE}
{van Straten}, W., {Demorest}, P., \& {Oslowski}, S. 2012, Astronomical
  Research and Technology, 9, 237

\bibitem[{Weber(1999)}]{weber1999pulsars}
Weber, F. 1999, Pulsars as Astrophysical Laboratories for Nuclear and Particle
  Physics, Series in High Energy Physics, Cosmology and Gravitation (Taylor \&
  Francis)

\bibitem[{Xie \& Zhang(2013)}]{slowgl3}
Xie, Y., \& Zhang, S.-N. 2013, The Astrophysical Journal, 778, 31

\bibitem[{{Yu} {$et~al$.}(2013){Yu}, {Manchester}, {Hobbs}, {Johnston},
  {Kaspi}, {Keith}, {Lyne}, {Qiao}, {Ravi}, {Sarkissian}, {Shannon}, \&
  {Xu}}]{Yu2013}
{Yu}, M., {Manchester}, R.~N., {Hobbs}, G., {$et~al$.} 2013, MNRAS, 429, 688

\bibitem[{Zhou {$et~al$.}(2019)Zhou, Zhou, Zhang, Liu, Liu, Zhang, Feng, Zhu,
  \& Wu}]{Zhou2019}
Zhou, S.~Q., Zhou, A.~A., Zhang, J., {$et~al$.} 2019, Astrophysics and Space
  Science, 364, 173

\end{thebibliography}

\end{document}